# Do We Preach What We Practice?
## Investigating the Practical Relevance of Requirements Engineering Syllabi – The IREB Case

Daniel Méndez Fernández[1], Xavier Franch[2], Norbert Seyff[3], Michael Felderer[4], Martin Glinz[5], Marcos Kalinowski[6], Andreas Vogelsang[7], Stefan Wagner[8], Stan Bühne[9], Kim Lauenroth[10]

[1] Technical University of Munich, Germany
[2] Universitat Politècnica de Catalunya, Spain
[3] University of Applied Sciences and Arts Northwestern Switzerland
[4] University of Innsbruck, Austria
[5] University of Zurich, Switzerland
[6] Pontifical Catholic University of Rio de Janeiro, Brazil
[7] Technical University of Berlin, Germany
[8] University of Stuttgart, Germany
[9] Seven Principles AG, Germany
[10] adesso AG, Germany

**Abstract.** Nowadays, there exist a plethora of different educational syllabi for Requirements Engineering (RE), all aiming at incorporating practically relevant educational units (EUs). Many of these syllabi are based, in one way or the other, on the syllabi provided by the International Requirements Engineering Board (IREB), a non-profit organisation devoted to standardised certification programs for RE. IREB syllabi are developed by RE experts and are, thus, based on the assumption that they address topics of practical relevance. However, little is known about to what extent practitioners actually perceive those contents as useful. We have started a study to investigate the relevance of the EUs included in the IREB Foundation Level certification programme. In the first phase reported in this paper, we have surveyed practitioners mainly from DACH countries (Germany, Austria and Switzerland) who participated in the IREB certification. Later phases will widen the scope both by including other countries and by not requiring IREB-certified participants. The results shall foster a critical reflection on the practical relevance of EUs built upon the de-facto standard syllabus of IREB.

**Keywords:** Survey Research, Requirements Engineering Education, Certification Programs.

## 1 Introduction

Requirements Engineering (RE) plays a vital role for successful software and systems development projects [1] and, thus, training and teaching relevant RE knowledge to practitioners is of utmost importance. The non-profit organisation IREB (International

Requirements Engineering Board, www.ireb.org) has developed standardised syllabi at different levels for the individual certification of professionals in RE. These syllabi are created by RE representatives from academia, industry, and consulting. The Foundation Level (FL) syllabus includes different RE educational units (EUs), which, according to IREB, cover essential knowledge areas and are used as a baseline for the certification to become a certified professional for RE. IREB training providers rely on the IREB syllabus to develop course material to train candidates for the certification exams.

As of today, more than 40,000 candidates all over the world have passed the IREB FL exam. IREB is thus becoming the de-facto standard when it comes to standardised education for RE. However, the IREB certification programme, same as other educational RE curricula built upon the IREB syllabi, are all based on the core assumption that the IREB FL syllabus covers practically relevant topics. Yet, there is limited knowledge about the extent to which practitioners actually perceive the content as useful and relevant in their work.

In order to better understand the relevance of the topics included in the IREB FL syllabus, we have designed a study with three different goals: (1) *how useful* practitioners consider its content for their daily work, (2) *how relevant* they perceive it for their personal knowledge and interest, and (3) what do they *miss*. This paper reports on a first iteration of this study, in which we present results gathered from practitioners in the DACH area (Germany, Austria and Switzerland) taking the IREB FL exam. We plan future iterations in which other regions (e.g., Latin America, Asia, …) are targeted, as well as RE practitioners in general, regardless whether they are certified by IREB or not. One hope we associate with this article is to motivate other researchers to join our endeavour. As an ultimate goal, this investigation allows us to critically reflect on the IREB FL syllabus, but also on other syllabi which are built upon it.

## 2       Fundamentals and Related Work

The IREB Foundation Level aims at providing the core knowledge of RE. The syllabus consists of nine educational units, see Table 1. The first version of the syllabus was developed by a group of RE researchers and practitioners in 2007 and since then, it has been revisited and extended. This paper is based on the current version of the syllabus 3 (Version 2.2). The goal of the syllabus has always been to define a basic RE body of knowledge that each RE practitioner should know. Therefore, it is important to assess the syllabus' content against the needs and demands of RE practitioners.

Besides our own work, there are other studies that aim at increasing the practical relevance of academic RE contributions in the long run. The NaPiRE initiative [1], for instance, constitutes a family of surveys on contemporary problems, causes, and effects in practice. Franch et al. [4] report on an ongoing survey-based study where practitioners assess the perceived relevance of RE research papers. Fricker et al. [5] surveyed several industrial RE projects to assess the current state of methods, tools, and techniques in RE practice. To the best of our knowledge, our research is the first aiming at the direct evaluation of current RE teaching syllabi from the perspective of practitioners' needs and expectations.

4**Table 1.** Educational Units (EU) in IREB FL.

| EU1 | Introduction and Foundations |
|---|---|
| EU2 | System and System Context |
| EU3 | Requirements Elicitation |
| EU4 | Requirements Documentation |
| EU5 | Documentation of Requirements using Natural Language |
| EU6 | Model-based Documentation of Requirements |
| EU7 | Requirements Validation and Negotiation |
| EU8 | Requirements Management |
| EU9 | Tool Support |

## 3 Study Design

In the following, we describe the overall design of our survey conducted in the IREB context. Details can also be taken from the disclosed replication package [2] that includes the export of raw data and analysis material as well as the full questionnaire.

### 3.1 Research Objectives and Questions

Our overall goal of the study is to critically reflect on the practical relevance of today's teaching material in Requirements Engineering and the extent to which it reflects practical needs. To this end, in this paper we aim at investigating what the perception of practitioners is on the usefulness and relevance of the IREB Foundation Level training material, i.e. the EUs of the syllabus, considering the perspective of practitioners taking part in the certification. We formulate the following research questions:

**RQ1.** How *useful* do practitioners perceive the content of RE educational material for their RE-related daily work?

**RQ2** How *relevant* is the content of RE educational material with respect to practitioners' knowledge and interests?

**RQ3** What topics do practitioners *miss*?

### 3.2 Instruments

To survey RE practitioners in the IREB certification programme context, we designed an instrument that captures various questions along two major parts:

- Demographics such as background and experience.
- Opinions, such as on relevance, on the educational units of the IREB FL syllabus.

Table 2 summarises the most relevant questions presented to practitioners in the survey as per selection with respect to our research questions. The full questionnaire, including details on the responses, can be found in our replication package [2].

**Table 2.** Summary of the questionnaire.

| Block *General questions* | |
|---|---|
| Q2.1 | What is your age? |
| Q2.2 | What is your educational background? |
| Q2.3 | What is your practical experience in requirements engineering? |
| Q2.4 | What kind of certification do you already have? |
| Q2.5 | How did you become aware of the IREB certification program? |
| Q2.6 | What kind of tools and methods do you use in your daily work? |
| Q2.8 | Do you have a CPRE FL certification? |
| Q2.9 | Are you interested in CPRE certifications? |
| Q2.11 | In which certifications are you interested? |
| Block *About the FL certification* | |
| Q3.1 | Since when (year) do you have your CPRE Foundation Level certification? |
| Q3.2 | Why did you choose CPRE Foundation Level? |
| Q3.4 | In which country did you get your CPRE Foundation Level certification? |
| Block *Assessment of FL syllabus* | |
| Q5.1 | Rate the different Educational Units (EUs) of the syllabus regarding their content (Note: for each of the 9 EUs, respondents rated: too generic – detailed enough – too many details – don't remember) |
| Q5.2 | How do you rate the overall structure of the syllabus? (Note: respondents rated from 1 –extremely good– to 7 –extremely bad–) |
| Q5.3 | How relevant are the following EUs for your personal knowledge and interest? (Note: for each of the 9 EUs, respondents rated from 1 –extremely interesting– to 5 –not interesting at all–) |
| Q5.4 | How useful are the following EUs in your requirements engineering related daily work? (Note: for each of the 9 EUs, respondents rated from 1 –extremely important– to 5 –not important at all–) |
| Q5.5 | What do you miss in the CPRE Foundation certification for your practical work? (Note: respondents rated four concepts –guidance in specific domains, methods and tools for the daily work, templates for my work, process focus rather than just providing a tool set– with either 1 –I want more–, 2 –It is okay as it is– or 3 –I want less–) |

### 3.3 Data Collection

We aim at answering our research questions via survey research. The IREB environment provides the context for the survey. IREB EUs for the FL certification represent the study object. The practitioners asked to participate in the study also stem from the same IREB environment.

The survey was implemented using Qualtrics Survey Software and invitations have been sent out as an newsletter sent via email to approximately 40.000 registered IREB practitioners with the overall purpose of receiving **feedback for a revision of the handbook (still ongoing)**. The data collection was conducted from April to August 2017. Prior to the survey, a pilot has been performed with 5 practitioners. To increase the motivation of the participants, a fitness smartwatch worth 250€ was raffled among all participants who completed the survey. Our respondents come mainly from the DACH area, i.e. Germany (D), Austria (A), and Switzerland (CH).

### 3.4 Data Analysis

Given the scope of the questions, we reduce the data analysis to basic descriptive statistics, including blocking of results according to chosen context factors such as experience. Qualitative data was analysed manually without an elaborate coding procedure due to the small number of answers.

## 4 Study Results

The theoretical population for the survey consists of practitioners taking the IREB FL exam (hereafter, "practitioners" for short). The survey received in total 137 responses. We analysed them and filtered out those that were incomplete leaving us with 108 practitioners' responses. Please note that not all survey questions were mandatory and the number of responses for single research questions varies slightly.

Looking at the age of the practitioners who participated in the survey, the majority of the respondents are fairly distributed between the ages 25-34 years (41), 35-44 years (33), and 45-54 years (26). Half of the sample (54) holds a professional degree. We measure the professional work experience of the respondents in a simplified manner via working years: 53 of the respondents have less than five years of experience, 39 between 5 and 15 years, and 16 have more than 15 years of experience. The certifications of our respondents are primarily in Requirements Engineering, Project Management, and Testing with 82, 50, and 27 respondents, respectively.

In the following, we report on the study results structured according to the research questions.

### 4.1 Usefulness for practitioners' daily work (RQ 1)

To analyse the usefulness of RE educational material, we asked our respondents to rate the individual IREB EUs according to their importance for their daily work using a Likert scale.

Table 3 gives an overview of the ratings our respondents gave to the individual EUs. All EUs in the list, except for EU1 (*Introduction and Foundations*) and EU9 (*Tool Support*), have been perceived as extremely important, very important or moderately important by more than 75% of the practitioners participating in the survey. For EU9, this percentage dropped to 53.41% and it dropped even to 50% for EU1.

We blocked those results by the respondents' working experience (see also the spreadsheets in the replication package [2] for a more detailed view) and we observed a trend in the perception: more experienced respondents seem to value the foundations more than less experienced ones. Indeed, 66.67% of the most experienced respondents (more than 15 years of experience) consider this EU extremely or very important, while for the least experienced ones (less than 5 years of experience) this number drops to 33.33%, while 40% of them still consider the EU to be moderately important. Our interpretation to these results is that experienced practitioners are aware of the complexities of RE and recognize the value of having an initial EU providing solid foundations for the rest of the syllabus.

As for EU9 (*Tool Support*), we couldn't find any trend in relation to the experience. One explanation might be that the content of this EU might not match the reality that practitioners face in their projects. To corroborate this argument, we point to our observation that the survey question rating the actually trained contents of these EUs (Q5.1) informs that 30% of the respondents rate EU9's content as too generic, being the worst ranked EU at this respect.

It is also worth pointing to the documentation of requirements, which spreads along three EUs: *Requirements Documentation* (EU4), *Documentation of Requirements in Natural Language* (EU5), and *Model-based Documentation* (EU6). We consider two of our observations to be relevant. First, the general EU4 is perceived as more useful than the two detailed ones (EU5 and EU6). As a slight nuance to this result, blocking the results by the degree of experience shows that more experienced practitioners tend to rank the two detailed EUs more important and the percentage of practitioners who rate the EU on natural language requirements as extremely important raises from 37.78% for less experienced respondents up to 67.74% for medium and even to 75% for very experienced ones. The model-based EU follows a very similar trend (increasing in the agreement from 35.55% to 56.25% and then to 58.33%). Our only explanation for that phenomenon is that the more experienced practitioners are, the more they tend to get specialised and to value more advanced techniques.

We have also analysed possible trends when blocking the results according to other factors. We observed, for instance, that respondents with a *Project Management* certificate tend to generally give a higher rating on the usefulness of the EUs than those with a different RE certificate (like REQB) or a testing certificate (e.g., ISTQB), with *Architecture* and especially *Business Process* certificates yielding the lowest ratings. The differences are even higher if we focus on the core EUs that deal with specific RE phases (EU3 to EU8). As the only but rather clear exception, the usefulness of model-based documentation (EU6) was rated higher by those having these last two certificates, being, in fact, the only core EU in which this higher rating was given. One explanation we have for this is that both software architects and process modellers value the use of models more. Concerning the other EUs, the rating of tooling (EU9) does again not align well with the rating of the core EUs. Respondents with architecture or testing certifications rate this EU with higher usefulness than the rest of certifications. We argue that the technical nature of such certifications provide respondents with a higher perception on the importance of tool support.

Finally, we found a potential relationship among the perceived usefulness and the motivations our respondents have for obtaining their certificates. Practitioners who stated to participating in the certification for personal interest give a higher rating than those participating because it was requested by their employers. This difference gets clearer if we consider the three highest ratings: in each EU, individuals who participated in the course motivated by personal interest, tend to rate the EU better than the others.

**Table 3.** Perceived usefulness of the IREB Educational Units (EUs) from a practitioners' perspective

| # | Field | Extremely important | | Very important | | Moderately important | | Slightly important | | Not at all important | | Total |
|---|---|---|---|---|---|---|---|---|---|---|---|---|
| 1 | EU1 - Introduction and Foundations | 5.68% | 5 | 15.91% | 14 | 28.41% | 25 | 28.41% | 25 | 21.59% | 19 | 88 |
| 2 | EU2 - System and System Context | 15.73% | 14 | 29.21% | 26 | 32.58% | 29 | 15.73% | 14 | 6.74% | 6 | 89 |
| 3 | EU3 - Requirements Elicitation | 19.10% | 17 | 39.33% | 35 | 26.97% | 24 | 10.11% | 9 | 4.49% | 4 | 89 |
| 4 | EU4 - Requirements Documentation | 24.72% | 22 | 43.82% | 39 | 21.35% | 19 | 8.99% | 8 | 1.12% | 1 | 89 |
| 5 | EU5 - Documentation of Requirements in Natural Language | 21.59% | 19 | 31.82% | 28 | 27.27% | 24 | 14.77% | 13 | 4.55% | 4 | 88 |
| 6 | EU6 - Model-based Documentation | 14.61% | 13 | 31.46% | 28 | 30.34% | 27 | 20.22% | 18 | 3.37% | 3 | 89 |
| 7 | EU7 - Requirements validation and negotiation | 12.36% | 11 | 25.84% | 23 | 39.33% | 35 | 17.98% | 16 | 4.49% | 4 | 89 |
| 8 | EU8 - Requirements Management | 20.22% | 18 | 26.97% | 24 | 30.34% | 27 | 14.61% | 13 | 7.87% | 7 | 89 |
| 9 | EU9 - Tool Support | 6.82% | 6 | 14.77% | 13 | 31.82% | 28 | 30.68% | 27 | 15.91% | 14 | 88 |

**Table 4.** Relevance to Practitioners' Interest of the IREB Educational Units (EUs)

| # | Field | Extremely interesting | | Very interesting | | Moderately interesting | | Slightly interesting | | Not interesting at all | | Total |
|---|---|---|---|---|---|---|---|---|---|---|---|---|
| 1 | EU1 - Introduction and Foundations | 3.41% | 3 | 28.41% | 25 | 37.50% | 33 | 22.73% | 20 | 7.95% | 7 | 88 |
| 2 | EU2 - System and System Context | 21.11% | 19 | 45.56% | 41 | 25.56% | 23 | 4.44% | 4 | 3.33% | 3 | 90 |
| 3 | EU3 - Requirements Elicitation | 20.22% | 18 | 52.81% | 47 | 19.10% | 17 | 6.74% | 6 | 1.12% | 1 | 89 |
| 4 | EU4 - Requirements Documentation | 25.56% | 23 | 37.78% | 34 | 30.00% | 27 | 5.56% | 5 | 1.11% | 1 | 90 |
| 5 | EU5 - Documentation of Requirements in Natural Language | 22.22% | 20 | 26.67% | 24 | 30.00% | 27 | 15.56% | 14 | 5.56% | 5 | 90 |
| 6 | EU6 - Model-based Documentation | 27.78% | 25 | 43.33% | 39 | 22.22% | 20 | 5.56% | 5 | 1.11% | 1 | 90 |
| 7 | EU7 - Requirements validation and negotiation | 18.89% | 17 | 46.67% | 42 | 24.44% | 22 | 7.78% | 7 | 2.22% | 2 | 90 |
| 8 | EU8 - Requirements Management | 22.22% | 20 | 45.56% | 41 | 24.44% | 22 | 4.44% | 4 | 3.33% | 3 | 90 |
| 9 | EU9 - Tool Support | 10.00% | 9 | 20.00% | 18 | 35.56% | 32 | 28.89% | 26 | 5.56% | 5 | 90 |

### 4.2 Relevance to practitioners' personal interests (RQ 2)

We asked the participants to rate the relevance of the individual EUs according to their personal knowledge and interest (see Table 4). The relevance of the 9 EUs composing the IREB FL syllabus was rated overall high. This means that for 6 out of the 9 EUs, more than 90% of the respondents gave a positive rating (at least moderately interesting). *Requirements Elicitation* (EU3) and *Model-based Documentation* (EU6) excel with more than 70% finding the topics at least highly interesting. Three EUs were reported with lower relevance. For *Documentation of Requirements in Natural Language* (EU5), 48.89% reported an extremely high or high relevance which went down to 31,82% for *Introduction and Foundations* (EU1) and 30.00% for *Tool Support* (EU9). For EU5, 21,1% reported a slight or no relevance at all, which raised to 30.68% for EU1 and even to 34.44% for EU9. The relatively low interest in these two last EUs is in tune with the results investigating their usefulness. In contrast, we observed a low interest in EU5, although it was ranked the 3rd most useful topic for practitioners' daily work (see Section 4.1).

When blocking our results again according to the experience of our respondents, the ones having an experience of between 5 and 10 years (32 participants) seemed to be more positive than the ones having less or more experience. On average, 25.72% of practitioners having between 5 and 15 years of experience rated the EUs as extremely relevant. This percentage drops to 16.33% for participants with an experience of less than 5 years (45 subjects), and down to 11.96% for practitioners having more than 15 years (12 respondents). Again, highly experienced participants rated EU9 lower. We couldn't find any trends when blocking the results according to any other criteria.

### 4.3 Missing Topics

To identify potentially missing topics, we asked the participants directly for 4 different categories: (1) guidance for specific domains, (2), methods and tools for their daily work, (3) templates for the work, and (4) overall process and approaches. The responses provided a rather mixed picture, in general seeming happy with the current extent of the content or wanting more, in particular for methods and tools and templates for their work (i.e. apparently a more practical focus). We blocked the results again according to the experience. The only trend that was apparent to us was that the less experienced our respondents are, the less they seem to be interested in guidance for specific domains. To investigate this topic further, we formulated also an open free text question. We received in total 42 answers and blocked the results again according to the practical experience. Very experienced practitioners with more than 15 years of experience (but with a rather low N=9) asked for more consideration of RE in the larger software engineering context as well as for advanced techniques to face specific project situations. This is reflected in statements like "[...] interview techniques (like minimum 7 W's for journalists), no grammar lectures as this is ground school stuff and does not help with reasonable content.", or "The syllabus should put a strong emphasis on creative finding and designing of solutions. In daily work it is expected from a requirements engineer to find and suggest innovative solutions not just to elicit and document requirements [...]."

The remaining respondents (N=33) stated either that nothing would be missing (12) or that they would miss more practical yet domain-independent guidance: "I miss a concrete approach, how the process might look in a real project [...]"; "Where to start? How to plan the process of RE? [...]"; "Templates, tools, practices"; "More real-life examples".

## 5      Discussion

Both the usefulness (RQ1) and the relevance (RQ2) of EU1 (Introduction and Foundations) and EU9 (Tool Support) are rated considerably lower than the other seven EUs. For EU9, this result is not surprising: 30 percent of the respondents consider this EU to be too generic (cf. Sect. 4.1). An analysis of the contents of EU9 confirms this finding. We were surprised, however, by the low rating of EU1, which covers foundational material. It seems that particularly the inexperienced respondents rated this EU low. Due to its introductory nature, the direct impact of EU1 on daily work might be difficult to perceive for inexperienced practitioners. Training providers should take this into account when deciding how to teach the material of EU1.

With respect to usefulness in practitioners' daily work, the EUs on requirements elicitation, documentation, and documentation in natural language are ranked highest. This is in line with the particular importance of these topics in today's industrial practice. Interestingly, the relevance of EU5 on documentation in natural language is rated lower than its usefulness. It seems that practitioners acknowledge the importance of natural language for expressing requirements, but do not find particular interest in the topic.

Concerning RQ3, our respondents did not identify major gaps in the syllabus. Experienced respondents seem to miss advanced techniques for dealing with specific project situations. However, this might go beyond the scope of a FL syllabus. Less experienced respondents tend to favour general-purpose processes rather than domain-specific or advanced techniques. We suggest to investigate whether inexperienced people would benefit if the syllabus had a stronger focus on general concepts, terms, and activities of requirements documentation and elicitation.

With respect to improvement potential of the syllabus, EUs 9 and 1 should be looked at first. The next candidates that could be better aligned with practical needs are (in this order) EUs 2, 7, 6, and 8. EU5 should be examined with respect to its rather low perceived relevance.

A final conclusion we can draw is that, in general, RE educational material could be better aligned with practical needs when it comes to specialised topics. Especially rather inexperienced students need to understand the characteristics of real-life RE situations and the high sensitivity of the various approaches and practices we have so far in RE to their context. In other words: a key take-away in the foundation of RE is that there is no such thing as a universal process. To especially drive home the complexity of real-life RE settings, educators thus need to rely on real-life examples while still aiming to provide a big picture (RQ3).

Our results indicate that the current IREB Foundation Level syllabus actually addresses the needs of its target audience, i.e., professionals having a limited amount of experience. Furthermore, experienced professionals also see value in such foundational topics. We believe that experienced professionals benefit from receiving holistic views on general, foundational terminology and practices which go beyond their daily work and provide them a broad perspective on the state of the art in RE.

## 6 Conclusions

In this paper, we report on a survey to explore the perceived usefulness and relevance of RE educational material, which we exemplarily analysed based on the educational units provided in the IREB Foundation Level syllabus. This survey has addressed the population of RE practitioners certified by IREB in DACH countries. Our findings indicate that this syllabus fits the needs of both inexperienced and experienced professionals. However, the results also provide indicators on how to increase the practical usefulness and relevance. At the same time, we postulate that usefulness for practice should not be the only quality criterion of a syllabus, especially in university or college-level education. This is also known as the RE education dilemma [7]).

Furthermore, there are still several limitations inherent to our study. First, some respondents might still have had problems distinguishing between the relevance for their personal knowledge and interest and the usefulness to their daily work. A richer investigation of such a differentiation would need, however, different forms of investigation that allows to better understand their preferences, motivations, and working context. Yet, the picture our results depict already shows some differences in the ratings which strengthens our confidence in the suitability of the used instrument. Hence, we believe that our survey provides already a first useful basis for an investigation of relevance and usefulness. Second, the response rate is very low. Therefore, our next step is to conduct other iterations using additional channels other than IREB. Generalising from our findings in the context of IREB to further educational material is dependent on the exact contents educators decide to deliver, and it also depends on the (pedagogical) way the content is exactly taught, which is out of the scope of the IREB syllabus. To provide richer conclusions, we also need further analyses. These are in scope of current work. We are currently analysing more context factors and we are working on a more elaborate statistical analysis considering also a complementary survey that captures the perspective of IREB training providers. The latter shall help grasping the perspective of practitioners and IREB training providers alike on the same aspects, thus allowing to delineate potential misconceptions. One hope we associate with this paper is also to motivate other researchers and educators to critically reflect on the practical relevance of RE educational material and to join us in the endeavour of exploring this highly diverse and sensitive topic.